\documentclass[twocolumn,showpacs,prc,floatfix]{revtex4-1}
\usepackage{graphicx}
\usepackage[dvips,usenames]{color}
\usepackage{amsmath,amssymb}
\usepackage{bm}
\newcommand{\bs}{\begin{sloppypar}} \newcommand{\es}{\end{sloppypar}}
\def\beq{\begin{eqnarray}} \def\eeq{\end{eqnarray}}
\def\beqstar{\begin{eqnarray*}} \def\eeqstar{\end{eqnarray*}}
\newcommand{\bal}{\begin{align}}
\newcommand{\eal}{\end{align}}
\newcommand{\beqe}{\begin{equation}} \newcommand{\eeqe}{\end{equation}}
\newcommand{\p}[1]{(\ref{#1})}

\begin {document}
\title{Phase transition to the state with nonzero average helicity in dense neutron matter
 }
\author{ A. A. Isayev}
\email{isayev@kipt.kharkov.ua}
 \affiliation{Kharkov Institute of
Physics and Technology, Academicheskaya Street 1,
 Kharkov, 61108, Ukraine
\\
Kharkov National University, Svobody Sq., 4, Kharkov, 61077, Ukraine
 }
  \author{J. Yang}
 \email{jyang@ewha.ac.kr}
 \affiliation{Department  of Physics and the Institute for the Early Universe,
 \\
Ewha Womans University, Seoul 120-750, Korea
}
\begin{abstract}
The possibility of the appearance of the states with a nonzero
average helicity in neutron matter is studied in the model with the
Skyrme effective interaction.   By providing the analysis of the
self-consistent equations  at zero temperature, it is shown that
neutron matter with the Skyrme BSk18 effective force undergoes at
high densities a phase transition to the state in which the
degeneracy with respect to helicity of neutrons is spontaneously
removed.
\end{abstract}
\pacs{21.65.Cd, 21.10.Hw, 26.60.-c, 21.30.Fe} \keywords{neutron
matter, Skyrme interaction, helicity, parity violation} \maketitle

The issue of spontaneous appearance of spin polarized states in
nuclear matter is a topic of a great current interest due to
relevance in astrophysics. In particular, the scenarios of supernova
explosion and cooling of neutron stars are essentially different,
depending on whether nuclear matter is spin polarized or not.    On
the one hand, the models with the Skyrme  effective nucleon-nucleon
(NN) interaction predict the occurrence of spontaneous spin
instability in nuclear matter at densities in the range from
$\varrho_0$ to $4\varrho_0$ for different parametrizations of the NN
potential~\cite{R}-\cite{IY} ($\varrho_0 \simeq 0.16\,{\rm fm}^{-3}$
is the nuclear saturation density). On the other hand, for the
models with the realistic NN interaction, no sign of spontaneous
spin instability  has been found so far at any isospin asymmetry up
to densities well above $\varrho_0$~\cite{PGS}-\cite{BB}. In order
to reconcile two different approaches, based on the effective and
realistic NN interactions, recently a new parametrization of the
Skyrme interaction, BSk18, has been proposed~\cite{CGP}, aimed to
avoid the spin instability of nuclear matter at densities beyond the
nuclear saturation density. This is achieved by adding new
density-dependent terms to the standard Skyrme force.  The advantage
of the BSk18 parametrization is that it also preserves the
high-quality fits to the mass data obtained with the conventional
Skyrme force as well as it satisfactorily reproduces the results of
microscopic neutron matter calculations (equation of
state~\cite{FP}, $^1S_0$ pairing gap~\cite{CLS}). Hence, this Skyrme
parametrization has a good potentiality in the studies of various
neutron star properties~\cite{RMK}.

In terms of Landau Fermi liquid parameters, the ferromagnetic
instability of neutron matter is prevented by the requirement
$G_0>-1$, where $G_0$ is the zeroth coefficient in the expansion of
the dimensionless spin-spin interaction amplitude on the Legendre
polynomials, $G=\sum_l\,G_lP_l(\cos\theta)$~\cite{LL}. Although this
condition can hold true for all relevant densities, nevertheless,
the first coefficient $G_1$, with increasing density,  can become
large and negative, so that the condition $G_1<-3$ could be reached
in the high-density region of neutron matter. These conditions,
considered together, mean quite strong attractive interaction
between neutron spins in the triplet spin state and repulsive, or
weakly attractive interaction in the singlet spin state. As a
result, neutron matter becomes unstable at high densities, and a new
state characterized by a nonzero average helicity of neutrons,
$\langle\bm{\sigma p}^0\rangle\not=0$, is formed. Such a possibility
was first studied with respect to an electron liquid in metals  in
Ref.~\cite{AC} and, later, in the framework of a microscopic model,
in Ref.~\cite{CV}. Our primary goal here is to develop the proper
formalism for the description of the states with a nonzero average
helicity in neutron matter and to provide the corresponding analysis
of the self-consistent equations for the BSk18 Skyrme force.

 The nonsuperfluid states of neutron matter are described
  by the normal distribution function of neutrons $f_{\kappa_1\kappa_2}=\mbox{Tr}\,\varrho
  a^+_{\kappa_2}a_{\kappa_1}$, where
$\kappa\equiv({\bf{p}},\sigma)$, ${\bf p}$ is momentum, $\sigma$
is the projection of spin on the third axis, and $\varrho$ is the
density matrix of the system~\cite{I,IY}.   The self-consistent
matrix equation for determining the distribution function $f$
follows from the minimum condition of the thermodynamic
potential~\cite{AIP,IY3} and is
  \begin{eqnarray}
 f=\left\{\mbox{exp}(Y_0\varepsilon+
Y_4)+1\right\}^{-1}\equiv
\left\{\mbox{exp}(Y_0\xi)+1\right\}^{-1}.\label{2}\end{eqnarray}
Here the single particle energy $\varepsilon$ and quantity $Y_4$
are the matrices in the space of $\kappa$ variables, with
$Y_{4\kappa_1\kappa_2}=Y_{4}\delta_{\kappa_1\kappa_2}$, $Y_0=1/T$,
and $ Y_{4}=-\mu_0/T$  being
 the Lagrange multipliers, $\mu_0$ being the chemical
potential of  neutrons, and $T$  being the temperature.

For the BSk18 interaction, the ferromagnetic instability is
avoided by adding to the standard Skyrme force the additional
density-dependent terms. Nevertheless, although spontaneous spin
polarization is excluded at all densities relevant for neutron
stars, there is still the possibility, related to the spontaneous
appearance of the state with a nonzero average helicity
\begin{align}
\lambda\equiv\langle\bm{\sigma}\mathbf{p}^0\rangle, \;
\mathbf{p}^0=\mathbf{p}/p, \label{spir}\end{align} where
$\langle\ldots\rangle\equiv {\rm tr}\,f.../{\rm tr}\,f$, ${\rm
tr}\,...$ being the trace in the space of  $\kappa$ variables.
 For this state, the single
particle energy  of neutrons
reads 
\begin{align}
\varepsilon({\bf p})&= \varepsilon_{0}({\bf p})\sigma_0
-\Delta(\bm{p}){\bm{\sigma}}\bm{p}^0,\label{eps}
\end{align}
where $\sigma_i$ are the Pauli matrices in the spin space and
$2\Delta(\bm{p})$ is the energy splitting between the neutron
spectra with different helicities (positive and negative).  It
follows from Eqs.~\p{2} and \p{eps}, that the distribution
function of neutrons has the structure
\begin{align}
 f({\bf p})&=
f_{0}({\bf p})\sigma_0+f_{||}({\bf
p})\bm{\sigma}\bm{p}^0,\label{f}
\end{align}
 where
\begin{align} f_{0}=\frac{1}{2}\{n(\omega_{+})+n(\omega_{-})
\},\;
f_{||}=\frac{1}{2}\{n(\omega_{+})-n(\omega_{-})\}.\label{fpar}
 \end{align}  Here $n(\omega)=\{\exp(Y_0\omega)+1\}^{-1}$
 and
 \bal
\omega_{\pm}&=\varepsilon_{0}-\mu_{0}\mp\Delta.\end{align}

The quantity $\omega_{\pm}$, being the exponent in the Fermi
distribution function $n$, entering Eqs.~\p{fpar}, plays the role of
the quasiparticle spectrum. There are two branches of the
quasiparticle spectrum, corresponding to neutrons with definite
helicity,  $\bm{\sigma}\bm{p}^0=\pm1$.

Note that the distribution function $f_0$  should satisfy the
normalization condition
\begin{align} \frac{2}{\cal
V}\sum_{\bf p}f_{0}({\bf p})&=\varrho,\label{3.1}\end{align} where
$\varrho$ is the total density of
 neutron matter. The average helicity $\lambda$ plays the role of
 the order parameter of a phase transition to the state, in which the majority of neutron
 spins are oriented along, or opposite to their momenta. By calculating the traces in
 Eq.~\p{spir},
 it is easy to find
that
\begin{equation}
\lambda=\frac{\sum_{\bf p}f_{||}({\bf p})}{\sum_{\bf p}f_{0}({\bf
p})}. \label{lamb}\end{equation}

In order to get the self--consistent equations for the components of
the single particle energy, one has to set the energy functional of
the system, which reads~\cite{IY,AIP}
\bal E(f)&=E_0(f)+E_{int}(f),\label{enfunc} \\
E_0(f)&=2\sum\limits_{\bf p}^{}\underline{\varepsilon}_{\,0}({\bf
p})f_{0}({\bf
p}),\;\underline{\varepsilon}_{\,0}({\bf p})=\frac{{\bf p}^{\,2}}{2m_{0}},\nonumber\\
 E_{int}(f)&=\sum\limits_{\bf p}^{}\{\tilde\varepsilon_{0}({\bf p})f_{0}({\bf
p})+\tilde{{\varepsilon_i}}({\bf p}){f_i}({\bf p})\},\; \nonumber
\end{align}
where
\begin{align}\tilde\varepsilon_{0}({\bf p})&=\frac{1}{2\cal
V}\sum_{\bf q}U_0^n({\bf k})f_{0}({\bf
q}),\;{\bf k}=\frac{{\bf p}-{\bf q}}{2}, \label{flenergies}\\
\tilde{{\varepsilon}_i}({\bf p})&=\frac{1}{2\cal V}\sum_{\bf
q}U_1^n({\bf k}){f_i}({\bf q}),\; {f_i}({\bf q})=f_{||}({\bf
q})q_i^0.\label{epsvec}
\end{align}
Here  $\underline{\varepsilon}_{\,0}({\bf p})$ is the free single
particle spectrum, $m_0$ is the bare mass of a neutron, $U_0^n({\bf
k}), U_1^n({\bf k})$ are the normal Fermi liquid (FL) amplitudes,
and $\tilde\varepsilon_{0},\tilde{\varepsilon}_i$ are the FL
corrections to the free single particle spectrum.   Using equation
$\delta E={\rm tr}\, \varepsilon(f)\delta f$~\cite{LL}, we get the
self-consistent equations in the form \bal\xi_{0}({\bf
p})&=\underline{\varepsilon}_{\,0}({\bf
p})+\tilde\varepsilon_{0}({\bf p})-\mu_0,\; \label{ksi0}\\
{\xi_i}({\bf p})&\equiv -\Delta({\bf
p}){p_i}^0=\tilde{\varepsilon}_i({\bf p}).\label{14.2}
\end{align}

   To obtain
 numerical results, we  utilize the  BSk18 parametrization of the
 Skyrme interaction, developed in Ref.~\cite{CGP} and generalizing the
 conventional Skyrme parametrizations.  In the conventional case,
 the amplitude of NN interaction reads~\cite{VB} \bal\hat v({\bf p},{\bf
q})&=t_0(1+x_0P_\sigma)+\frac{1}{6}t_3(1+x_3P_\sigma)\varrho^\alpha
\label{49}\\&+\frac{1}{2\hbar^2} t_1(1+x_1P_\sigma)({\bf p}^2+{\bf
q}^2) +\frac{t_2}{\hbar^2}(1+x_2P_\sigma){\bf p}{\bf
q},\nonumber\end{align} where
$P_\sigma=(1+{{\boldsymbol\sigma_1\boldsymbol\sigma_2}})/2$ is the
spin exchange operator,  $t_i, x_i$ and $\alpha$ are some
phenomenological parameters specifying a given parametrization of
the Skyrme interaction. The Skyrme interaction suggested in
Ref.~\cite{CGP}  has the form \bal \hat v'({\bf p},{\bf q})&=\hat
v({\bf p},{\bf q})+ \frac{\varrho^\beta}{2\hbar^2}
t_4(1+x_4P_\sigma)({\bf p}^2+{\bf
q}^2)\label{comSk}\\&\quad+\frac{\varrho^\gamma}{\hbar^2}t_5(1+x_5P_\sigma){\bf
p}{\bf q}.\nonumber\end{align} In Eq.~\p{comSk},  two additional
terms are the density-dependent generalizations of the $t_1$ and
$t_2$ terms of the usual form.

The normal FL amplitudes $U_0,U_1$ can be expressed in terms of
the Skyrme
  force parameters. For conventional Skyrme force parametrizations,
their explicit expressions are given in Refs.~\cite{AIP,IY3}. As
follows from Eqs.~\p{49} and \p{comSk}, in order to obtain the
corresponding expressions for the generalized Skyrme
interaction~\p{comSk}, one should use the substitutions
\begin{align} t_1\rightarrow t_1+t_4\varrho^\beta,\;
t_1x_1\rightarrow
t_1x_1+t_4x_4\varrho^\beta,\\
t_2\rightarrow t_2+t_5\varrho^\gamma,\; t_2x_2\rightarrow
t_2x_2+t_5x_5\varrho^\gamma.\end{align}

  Therefore, the FL amplitudes are related to the parameters of the Skyrme
interaction~\p{comSk} by formulas \bal U_0^n({\bf
k})&=2t_0(1-x_0)+\frac{t_3}{3}\varrho^\alpha(1-x_3)
+\frac{2}{\hbar^2}[t_1(1-x_1)\label{101}\\&
+t_4(1-x_4)\varrho^\beta+3t_2(1+x_2)+
3t_5(1+x_5)\varrho^\gamma]{\bf k}^{2},
\nonumber\\
U_1^n({\bf
k})&=-2t_0(1-x_0)-\frac{t_3}{3}\varrho^\alpha(1-x_3)+\frac{2}{\hbar^2}[t_2(1+x_2)
\label{102}\\&\quad
+t_5(1+x_5)\varrho^\gamma-t_1(1-x_1)- t_4(1-x_4)\varrho^\beta]{\bf
k}^{2}\nonumber\\ &\quad\equiv a_n+b_n{\bf
k}^{2}.\nonumber\end{align}  It follows from Eqs.~\p{ksi0} and
\p{101} that
\begin{equation}
\xi_{0}=\frac{p^2}{2m_{n}}-\mu,\label{4.32}\end{equation} where
the effective neutron mass $m_{n}$  is defined by
 the formula
\bal
\frac{\hbar^2}{2m_{n}}&=\frac{\hbar^2}{2m_0}+\frac{\varrho}{8}
[t_1(1-x_1)+t_4(1-x_4)\varrho^\beta\label{181}\\&\quad+3t_2(1+x_2)+3t_5(1+x_5)\varrho^\gamma],
\nonumber\end{align} and the renormalized chemical potential $\mu$
should be determined from Eq.~\p{3.1}.

Taking into account the explicit form of the FL amplitude $U_1$ in
Eq.~\p{102}, solution of Eq.~\p{14.2} for the energy gap $\Delta$
should be sought in the form
\begin{equation} \Delta(p)=\frac{b_n}{4}\nu p,\label{delta}\end{equation}
where $\nu$ is some unknown quantity  satisfying the equation
\begin{equation}
\nu=\int_0^\infty
\frac{q^3}{6\pi^2\hbar^3}f_{||}(q)d\,q.\label{nu}
\end{equation}
This equation can be obtained from Eqs.~\p{epsvec},\p{14.2} after
passing from summation to integration,
$\frac{1}{V}\sum\ldots\rightarrow\int\,\frac{d^{\,3}q}{(2\pi\hbar)^3}\ldots$,
and performing then the  angle integration.

Thus, with account of Eqs.~\p{fpar}  for the distribution
functions $f$, we obtain the self--consistent equations \p{3.1}
and \p{nu} for the renormalized chemical potential $\mu$ and the
unknown $\nu$, determining the splitting $\Delta$ in the energy
spectrum~\p{eps} of neutrons with different helicities. Note that
the energy spectrum~\p{eps} is invariant under the time reversion
but not under the parity transformation. Hence, the state with
$\Delta\not=0$ is characterized by a spontaneously broken
$P$-symmetry.

\begin{figure}[tb]
\begin{center}
\includegraphics[width=7.6cm,keepaspectratio]{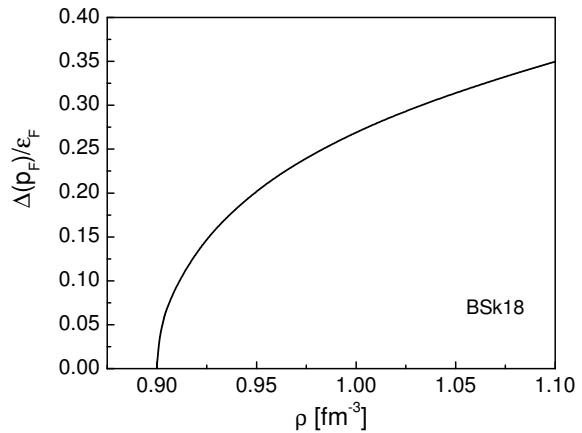}
\end{center}
\vspace{-2ex} \caption{The energy splitting $\Delta(p_F)$ between
the neutron spectra with  different helicities normalized to the
neutron Fermi energy as a function of density at zero temperature
for BSk18 interaction. } \label{fig1}\vspace{-0ex}
\end{figure}


Now we present the solutions of the self-consistent equations at
zero temperature for BSk18 Skyrme force~\cite{CGP}. Note that the
self-consistent equations have always the trivial solution
$\Delta=0$ (or $\nu=0$), corresponding to the normal neutron
matter. Besides, the self-consistent equations are invariant with
respect to  the change $\Delta\rightarrow-\Delta$, and, hence,
nontrivial solutions for $\Delta$ enter in a pair with the same
magnitude and opposite sign. The majority of neutrons will have
positive helicity, if $\Delta>0$, and  negative helicity, if
$\Delta<0$. After solving the self-consistent equations, the
average helicity $\lambda$, playing the role of the order
parameter of a phase transition,  can be obtained from
Eq.~\p{lamb}. According to Eq.~\p{lamb},  both signs of the
helicity  of the given  magnitude are possible because of the two
possible signs of the energy splitting $\Delta$. Note that in
order to preserve the realistic EoS of neutron matter obtained in
Ref.~\cite{FP}, the following constraints on the additional
parameters of the BSk18 parametrization were set
\begin{equation} \beta=\gamma,\; t_4(1-x_4)=-3t_5(1+x_5).
\end{equation}
Because of these constraints, the $t_4$ and $t_5$ terms cancel
completely in the FL amplitude $U_0$ and in the effective neutron
mass $m_n$, and only the FL amplitude $U_1$ is affected by the new
terms.

\begin{figure}[tb]
\begin{center}
\includegraphics[width=7.6cm,keepaspectratio]{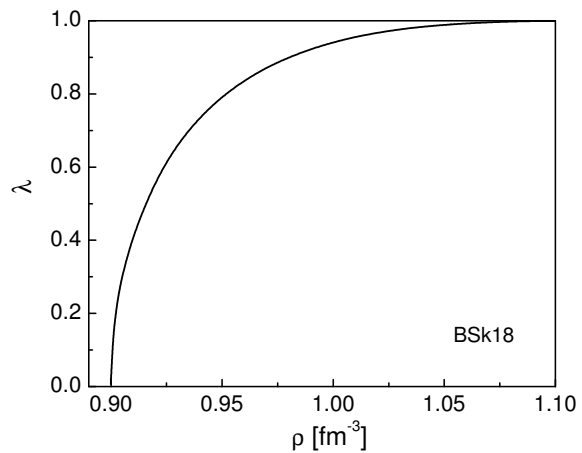}
\end{center}
\vspace{-2ex} \caption{Same as in Fig.~\ref{fig1} but for the
average helicity of neutron matter. } \label{fig2}\vspace{-0ex}
\end{figure}

Fig.~\ref{fig1} shows the energy splitting $\Delta(p=\hbar k_F)$
between the branches of the neutron spectra with different
helicities normalized to the Fermi energy
$\varepsilon_F=\frac{\hbar^2k_F^2}{2m_n}$ of the normal neutron
matter as a function of density. A spontaneous phase transition to
the state with a nonzero helicity occurs at the critical density
$\varrho\approx5.67\,\varrho_0$ ($\varrho_0=0.1586\,{\rm fm}^{-3}$
for BSk18 force). The energy splitting continuously increases with
the density and becomes comparable with the neutron Fermi energy
$\varepsilon_F$. Note that only the  branch with the positive
energy gap is shown in Fig.~\ref{fig1} while the symmetric branch
($\Delta\rightarrow-\Delta$) with the negative energy gap is not
presented there.

Fig.~\ref{fig2} shows the average helicity of neutron matter as a
function of density obtained with the BSk18 Skyrme interaction.
The average helicity monotonously increases from zero till it is
saturated and reaches the value $\lambda=1$  at
$\varrho\approx6.93\,\varrho_0$. Beginning from that density, all
neutron spins will be aligned along their momenta (or opposite to
them for the branch with the negative helicity, not shown in
Fig.~\ref{fig2}).

\begin{figure}[tb]
\begin{center}
\includegraphics[width=7.6cm,keepaspectratio]{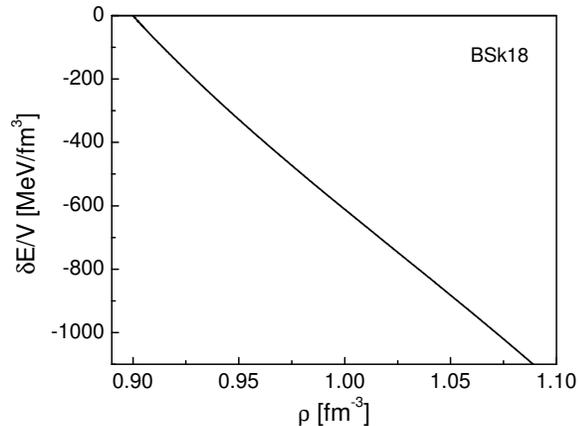}
\end{center}
\vspace{-3ex} \caption{Same as in Fig.~\ref{fig1} but for the
difference between the energy densities of the state with a
nonzero average helicity and the normal state ($\lambda=0$) of
neutron matter. } \label{fig3}\vspace{-0ex}
\end{figure}

In order to clarify whether the state with a nonzero average
helicity is thermodynamically preferable over the normal state of
neutron matter, we should compare the corresponding energies (at
zero temperature). Fig.~\ref{fig3} shows the difference between
the energy densities of the state with a nonzero average helicity
and the normal state of neutron matter. It is seen that for all
densities where nontrivial solutions (with $\Delta\not=0$) exist,
this difference is negative and, hence, the state with the
majority of neutron spins directed along (or opposite) to their
momenta is preferable at that density range.

In summary, we have considered the states with a spontaneous nonzero
average helicity in neutron matter with the BSk18 Skyrme NN
interaction, which are characterized by broken parity. The
self-consistent equations for the parameter, determining the energy
splitting between the neutron spectra  with different helicities,
and the effective chemical potential of neutrons have been obtained
and analyzed at zero temperature. It has been shown that the
self-consistent equations have solutions corresponding to a nonzero
average helicity beginning from the critical density
$\varrho\approx5.67\,\varrho_0$. Under increasing density, the
magnitude of the average helicity increases and is saturated at the
density $\varrho\approx6.93\,\varrho_0$, when all neutron spins are
aligned along ($\lambda=1$), or opposite ($\lambda=-1$) to their
momenta. The comparison of the respective energies at zero
temperature shows that the state with a nonzero average helicity is
preferable over the normal state at the densities beyond the
critical one. The possible existence of the state with a nonzero
average helicity in the dense core of a neutron star will affect the
neutrino opacities, and, hence, may be of importance for the
adequate description of the thermal evolution of pulsars. In this
respect, it is interesting to study also the impact of a strong
magnetic field, characteristic for pulsars (magnetars), on the
average helicity of dense neutron matter~\cite{IY4}.

J. Y. was supported by grant 2010-0011378 from Basic Science
Research Program through NRF of Korea funded by MEST and by grant
R32-2009-000-10130-0 from WCU project of MEST and NRF through Ewha
Womans University.



\end{document}